\documentclass{article}\usepackage[]{graphicx}\usepackage[]{color}
\makeatletter
\def\maxwidth{ %
  \ifdim\Gin@nat@width>\linewidth
    \linewidth
  \else
    \Gin@nat@width
  \fi
}
\makeatother

\definecolor{fgcolor}{rgb}{0.345, 0.345, 0.345}
\newcommand{\hlnum}[1]{\textcolor[rgb]{0.686,0.059,0.569}{#1}}%
\newcommand{\hlstr}[1]{\textcolor[rgb]{0.192,0.494,0.8}{#1}}%
\newcommand{\hlcom}[1]{\textcolor[rgb]{0.678,0.584,0.686}{\textit{#1}}}%
\newcommand{\hlopt}[1]{\textcolor[rgb]{0,0,0}{#1}}%
\newcommand{\hlstd}[1]{\textcolor[rgb]{0.345,0.345,0.345}{#1}}%
\newcommand{\hlkwb}[1]{\textcolor[rgb]{0.69,0.353,0.396}{#1}}%
\newcommand{\hlkwc}[1]{\textcolor[rgb]{0.333,0.667,0.333}{#1}}%
\newcommand{\hlkwd}[1]{\textcolor[rgb]{0.737,0.353,0.396}{\textbf{#1}}}%

\usepackage{framed}
\makeatletter
\newenvironment{kframe}{%
 \def\at@end@of@kframe{}%
 \ifinner\ifhmode%
  \def\at@end@of@kframe{\end{minipage}}%
  \begin{minipage}{\columnwidth}%
 \fi\fi%
 \def\FrameCommand##1{\hskip\@totalleftmargin \hskip-\fboxsep
 \colorbox{shadecolor}{##1}\hskip-\fboxsep
     \hskip-\linewidth \hskip-\@totalleftmargin \hskip\columnwidth}%
 \MakeFramed {\advance\hsize-\width
   \@totalleftmargin\z@ \linewidth\hsize
   \@setminipage}}%
 {\par\unskip\endMakeFramed%
 \at@end@of@kframe}
\makeatother

\definecolor{shadecolor}{rgb}{.97, .97, .97}
\definecolor{messagecolor}{rgb}{0, 0, 0}
\definecolor{warningcolor}{rgb}{1, 0, 1}
\definecolor{errorcolor}{rgb}{1, 0, 0}
\newenvironment{knitrout}{}{} 

\usepackage{alltt}
\usepackage{url}
\usepackage{natbib}
\usepackage{url}
\usepackage{hyperref}
\usepackage{amsmath,amsfonts}
\usepackage{fullpage}

\IfFileExists{upquote.sty}{\usepackage{upquote}}{}
\begin{document}
%
%
\title{ensr: R Package for Simultaneous Selection of Elastic Net Tuning Parameters}
\author{Peter E.\ DeWitt, Ph.D.\,$^{1}$,Tellen D. Bennett, MD, MS\,$^{2,3,4}$}
\date{%
  \small
  $^{1}$Department of Radiology, University of Colorado School of Medicine, Aurora, 80045, United States of America \\%
  $^{2}$Department of Pediatrics, Section of Pediatric Critical Care, University of Colorado School of Medicine, Aurora, 80045, United States of America \\%
  $^{3}$Adult and Child Consortium for Health Outcomes Research and Delivery Science, Children's Hospital Colorado, University of Colorado School of Medicine, Aurora, 80045, United States of America \\%
  $^{4}$Data Science to Patient Value, University of Colorado Anschutz Medical Campus, Aurora, 80045, United States of America \\[2ex]%
  \today
}




\maketitle

\abstract{
  {\bfseries Motivation:}\\
  Elastic net regression is a form of penalized regression that lies between
  ridge and least absolute shrinkage and selection operator (LASSO) regression.
  The elastic net penalty is a powerful tool for controlling the impact of correlated predictors and the
  overall complexity of generalized linear regression models. The elastic net
  penalty has two tuning parameters: $\lambda$ for the complexity and $\alpha$
  for the compromise between LASSO and ridge.  The R package
  provides efficient tools for fitting elastic net models and selecting $\lambda$
  for a given $\alpha.$ However, {\tt glmnet} does not simultaneously search the
  $\lambda-\alpha$ space for the optimal elastic net model.\\
  {\bfseries Results:}\\
  We built the R package {\tt ensr}, \emph{e}lastic \emph{n}et
  \emph{s}earche\emph{r}. {\tt ensr} extends the functionality of {\tt glmnet}
  to search the $\lambda-\alpha$ space and identify an optimal
  $\lambda-\alpha$ pair.\\
  {\bfseries Availability:}\\
  {\tt ensr} is available from the Comprehensive R Archive Network at
  \url{https://cran.r-project.org/package=ensr}.
\\
\textbf{Contact:} \href{peter.dewitt@ucdenver.edu}{peter.dewitt@ucdenver.edu}\\
}


\section*{Introduction}

Elastic net regression \citep{zou2005regularization} is a penalized linear
modeling approach that lies between ridge regression \citep{hoerl1970ridge},
and least absolute shrinkage and selection operator (LASSO) regression \citep{tibshirani1996regression}.
Ridge regression reduces the impact of collinearity on model parameters and LASSO
reduces the dimensionality of the support by shrinking some of the regression
coefficients to zero.  Elastic net does both of these.
Specifically, for a linear model of the form
\begin{equation}
  \label{eq:linear_model}
  E \left( \left. \boldsymbol{Y}  \right| \boldsymbol{X} = \boldsymbol{x}
  \right) = \beta_0 + \boldsymbol{x}^{\top} \boldsymbol{\beta},
\end{equation}
the elastic net estimates $\boldsymbol{\beta}$ by solving \citep{JSSv033i01}
\begin{equation}
  \label{eq:penalized_likelihood_part1}
  \min_{ \left(\beta_0, \boldsymbol{\beta} \right) \in \mathbb{R}^{p + 1} }
  \left[ \frac{1}{2N} \sum_{i = 1}^{N} \left(y_i - \beta_0 - x_i^{\top}
  \boldsymbol{\beta}\right)^2 + \lambda P_{\alpha} \left(\boldsymbol{\beta}
  \right) \right],
\end{equation}
where
\begin{equation}
  \label{eq:penalized_likelihood_part2}
  P_{\alpha} \left( \boldsymbol{\beta} \right) =
  \sum_{j = 1}^{p} \left[
    \frac{1}{2} \left(1 - \alpha \right) \beta_j^2 + \alpha \left| \beta_j
    \right| \right],
\end{equation}
$N$ is the sample size, and there are $p$ predictor variables.
$\lambda$ is a model complexity penalty that shrinks some $\beta_i$s to
zero. $\alpha$ defines whether the elastic net model is more similar to its
LASSO $(\alpha = 1)$ or ridge analogs $(\alpha = 0).$

The R package {\tt glmnet} \citep{JSSv033i01} provides tools for fitting
elastic net regression models. The {\tt glmnet} algorithms efficiently evaluate
multiple possible values of $\lambda$ when the user inputs a value of $\alpha$ (thereby
choosing ridge, elastic net, or LASSO regression). {\tt glmnet} can also
select $\lambda$ in an automated fashion using cross-validation. However, it lacks a
tool to simultaneously select values of $\alpha$ and $\lambda.$ Our package, {\tt ensr},
\emph{e}lastic \emph{n}et \emph{s}earche\emph{r}, provides a cross-validation
approach to select $\alpha$ and $\lambda$ simultaneously.

\begin{figure*}[t!]
  \includegraphics[width=\textwidth]{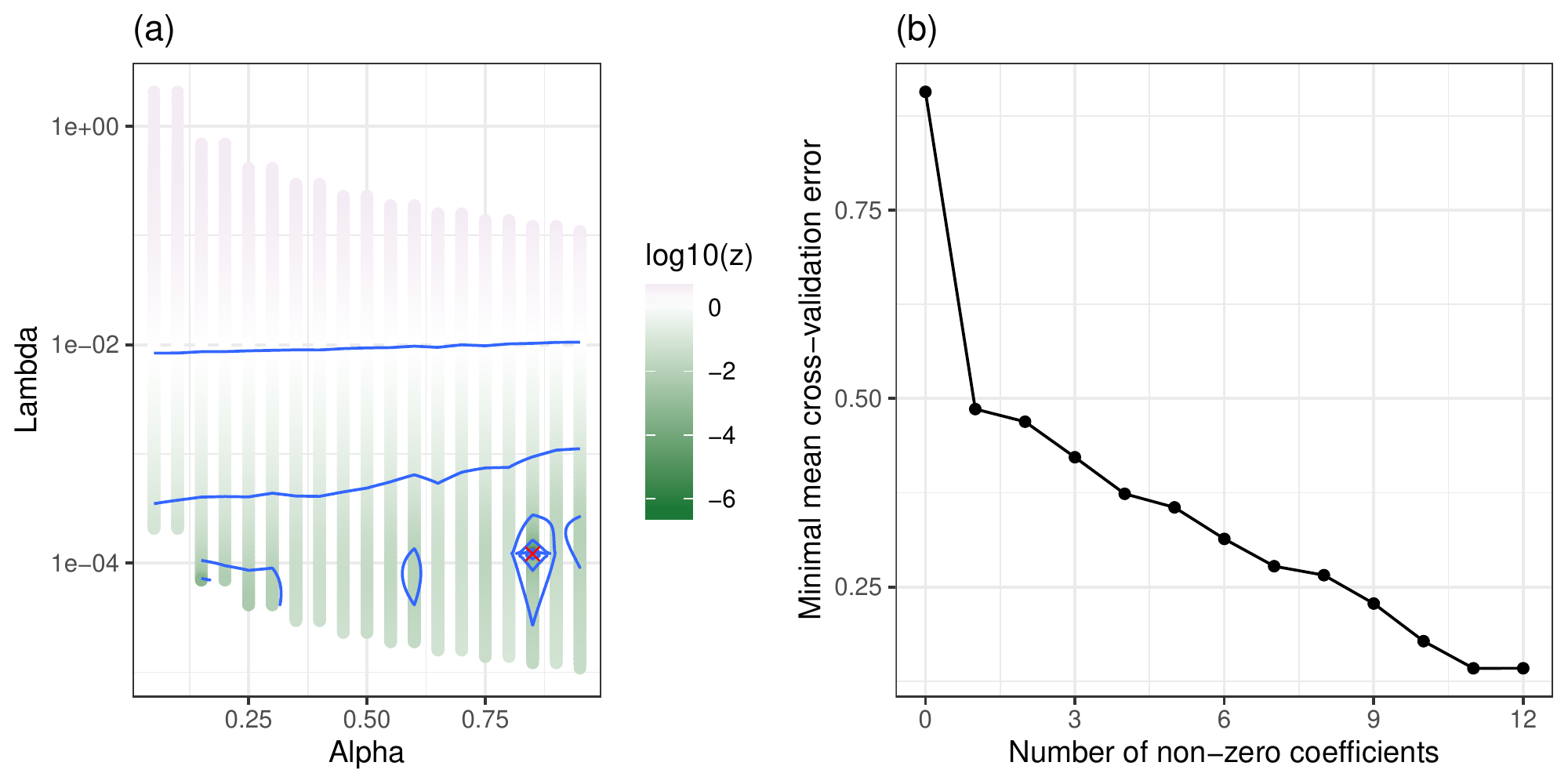}
  \caption{\label{fig:fig1} Plotting an {\tt ensr\_obj} generates two figures. The
  left panel (a) is a contour plot of $\lambda-\alpha$ space with the
  surface defined by $\log_{10}\left(Z\right),$ where $Z$ is the number of
  standard deviations above the global minimum observed cross-validation error.
  The overall minimum error is denoted with a red dot.
  The right panel (b) shows 
  the minimum cvm by the number of non-zero
  coefficients. This plot is provided to help users select parsimonious models
  with negligibly larger mean cross-validation error than the overall
  minimum.}
\end{figure*}

\section*{Usage}

The {\tt ensr} package is available from the Comprehensive R Archive Network
(CRAN) at \url{https://cran.r-project.org/package=ensr} and from
\url{https://github.com/dewittpe/ensr}.

After installation, load and attach the {\tt ensr} namespace in an active R
session. We encourage you to load and attach the {\tt data.table}
\citep{datatable} namespace as well.

Users interact with the eponymous function {\tt ensr}.
{\tt ensr::ensr} is an extension of {\tt glmnet::cv.glmnet}.
The two functions have the same API except that {\tt ensr} has one additional
argument, {\tt alphas}, to define a numeric vector of specific $\alpha$ values
to consider.  In contrast, {\tt cv.glmnet} will only consider a single value of
$\alpha.$

We included a synthetic example data set {\tt tbi} in the {\tt ensr} package based on \citet{bennett2017development}. The following example of {\tt ensr usage} aims to identify which of six procedure codes and six billing codes are
associated with particular injury type. We will model the odds of {\tt injury2}
as a function of the presence or absence of procedure codes {\tt pcode1, \ldots,
pcode6} and billing codes {\tt ncode1, \ldots, ncode6}. Additional examples using other
example data sets are available in the `ensr-examples' vignette.

As with {\tt glmnet::cv.glmnet}, the end user will need to construct matrices
for the response and the predictors. By default, {\tt ensr} will standardize all
predictors. In the below example, all predictors are binary, 0 (false) / 1 (true).
Therefore, standardization is not required. The sequence of {\tt alphas} defined below has been selected to
explicitly prevent either a pure ridge ($\alpha = 0$) or pure LASSO ($\alpha = 1$) model.

\begin{knitrout}
\definecolor{shadecolor}{rgb}{0.969, 0.969, 0.969}\color{fgcolor}\begin{kframe}
\begin{alltt}
\hlstd{ymat} \hlkwb{<-} \hlkwd{as.matrix}\hlstd{(tbi[, injury2])}
\hlstd{xmat} \hlkwb{<-} \hlkwd{as.matrix}\hlstd{(tbi[, pcode1}\hlopt{:}\hlstd{ncode6])}
\hlkwd{set.seed}\hlstd{(}\hlnum{2018}\hlstd{)}
\hlstd{ensr_obj} \hlkwb{<-}
  \hlkwd{ensr}\hlstd{(}\hlkwc{x} \hlstd{= xmat,}
       \hlkwc{y} \hlstd{= ymat,}
       \hlkwc{alphas} \hlstd{=} \hlkwd{seq}\hlstd{(}\hlnum{0.05}\hlstd{,} \hlnum{0.95}\hlstd{,} \hlkwc{length} \hlstd{=} \hlnum{10}\hlstd{),}
       \hlkwc{standardize} \hlstd{=} \hlnum{FALSE}\hlstd{,}
       \hlkwc{family} \hlstd{=} \hlstr{"binomial"}\hlstd{)}
\end{alltt}
\end{kframe}
\end{knitrout}


\section*{Output}

{\tt ensr} objects ({\tt ensr\_obj}) are lists of {\tt cv.glmnet} objects. Using an {\tt
ensr\_obj}, a user could take one of two approaches to select
a model: 1) Select the $\lambda-\alpha$ pair with the overall minimum
mean cross-validation error (minimum cvm), or 2) consider the trade-off between
the reduction in minimum cvm with the inclusion of additional non-zero
coefficients.

A summary of the {\tt ensr\_obj} gives the list index, {\tt l\_index}, for the
{\tt cv.glmnet} objects within {\tt ensr} object as well as the corresponding values
of $\lambda,$ $\alpha$, cvm, and the number of non-zero
coefficients, {\tt nzero}.

Selection of a preferable model can be automated or done with some analyst
input.  For example, consider the models with minimum cvm by number of non-zero
coefficients.

\begin{knitrout}
\definecolor{shadecolor}{rgb}{0.969, 0.969, 0.969}\color{fgcolor}\begin{kframe}
\begin{alltt}
\hlstd{by_nzero} \hlkwb{<-}
  \hlkwd{summary}\hlstd{(ensr_obj)[}\hlkwd{order}\hlstd{(cvm),}
                    \hlstd{.SD[cvm} \hlopt{==} \hlkwd{min}\hlstd{(cvm)],}
                    \hlkwc{by} \hlstd{= nzero]}
\hlstd{by_nzero[nzero} \hlopt{>} \hlnum{7}\hlstd{]}
\end{alltt}
\begin{verbatim}
##    nzero l_index    lambda    cvm alpha
## 1:    11      17 1.216e-04 0.1421  0.85
## 2:    12       3 6.959e-05 0.1422  0.15
## 3:    10      19 1.675e-03 0.1781  0.95
## 4:     9      19 3.618e-03 0.2280  0.95
## 5:     8      19 5.243e-03 0.2656  0.95
\end{verbatim}
\end{kframe}
\end{knitrout}

The model with the lowest cvm has twelve non-zero coefficients and could easily
be obtained using {\tt ensr::preferable}.

\begin{knitrout}
\definecolor{shadecolor}{rgb}{0.969, 0.969, 0.969}\color{fgcolor}\begin{kframe}
\begin{alltt}
\hlstd{pref_ensr_obj} \hlkwb{<-} \hlkwd{preferable}\hlstd{(ensr_obj)}
\end{alltt}
\end{kframe}
\end{knitrout}

There is very little difference in cvm between the model with twelve non-zero
coefficients and the model with eleven non-zero coefficients.  This can been
seen in the output of the summary call or from the graphic in
Figure~\ref{fig:fig1}.

\begin{knitrout}
\definecolor{shadecolor}{rgb}{0.969, 0.969, 0.969}\color{fgcolor}\begin{kframe}
\begin{alltt}
\hlkwd{plot}\hlstd{(ensr_obj,} \hlkwc{type} \hlstd{=} \hlnum{1}\hlstd{)} \hlcom{# Generate Fig. 1a}
\hlkwd{plot}\hlstd{(ensr_obj,} \hlkwc{type} \hlstd{=} \hlnum{2}\hlstd{)} \hlcom{# Generate Fig. 1b}
\end{alltt}
\end{kframe}
\end{knitrout}


\section*{Discussion}

As with other uses of cross-validation, selection of $\lambda$ and $\alpha$ is subject to the
number and membership of the folds. We recommend that {\tt ensr} users select
final models by bootstrapping cross-validation errors and perform sensitivity analyses
varying cross-validation fold membership.

Another R package, {\tt glmnetUtils} \citep{r-glmnetUtils}, also selects
$\lambda$ and $\alpha$ using cross-validation.
A major difference between {\tt ensr} and {\tt
glmnetUtils} is that {\tt ensr} evaluates a richer grid of $\lambda-\alpha$ pairs by default.
{\tt glmnetUtils} uses default {\tt glmnet} values of $\lambda$ for each $\alpha$ value.
This results in a unique set of $\lambda$ values for each $\alpha$. In contrast, {\tt
ensr} constructs a set of $\lambda$ values such that common values of
$\lambda$ are evaluated for multiple values of $\alpha.$ The {\tt ensr} contour plot
method gives the user a way to visualize the grid of $\lambda-\alpha$ pairs
evaluated.



\section*{Conclusion}

The {\tt ensr} R package extends the functionality of the {\tt glmnet} package
by providing users the ability to automate selection of
tuning parameters for elastic net regression models.


\section*{Funding}

This work was supported by \textit{Eunice Kennedy Shriver} National Institute of Child Health and Human
Development grant R03 HD094912 to TB.\vspace*{-12pt}

\bibliographystyle{natbib}
%
%


\end{document}